\documentclass[]{gIPE2e}

\usepackage{color}
\usepackage{multicol}
\usepackage{chngcntr}

\begin{document}
%\doi{10.1080/1741597YYxxxxxxxx}
% \issn{1741-5985}
%\issnp{1741-5977}

%\jvol{00} \jnum{00} \jyear{2012} \jmonth{xxxx}

\markboth{P. Tagade \& H.-L.~Choi}{SSP Bayesian}

%\articletype{Regular Paper}

\title{A Generalized Polynomial Chaos-Based Method for Efficient Bayesian Calibration of Uncertain Computational Models}

\author{Piyush. M. Tagade \& Han-Lim Choi$^{\rm a}$$^{\ast}$\thanks{$^\ast$Corresponding author. Email: hanlimc@kaist.ac.kr}\\ \vspace{6pt}  $^{\rm a}${\em{Division of Aerospace Engineering, KAIST, Daejeon 305-701, Republic of Korea}}
%\\\vspace{6pt}\received{v3.4 released May 2008}
}

\maketitle

\begin{abstract}
This paper addresses the Bayesian calibration of dynamic models with parametric and structural uncertainties, in particular where the uncertain parameters are unknown/poorly known spatio-temporally varying subsystem models.
Independent stationary Gaussian processes with uncertain hyper-parameters describe uncertainties of the model structure and parameters while  Karhunnen-Loeve expansion is adopted to spectrally represent these Gaussian processes.
The Karhunnen-Loeve expansion of a prior Gaussian process is projected on a generalized Polynomial Chaos basis, whereas intrusive Galerkin projection is utilized to calculate the associated coefficients of the simulator output.
Bayesian inference is used to update the prior probability distribution of the generalized Polynomial Chaos basis, which along with the chaos expansion coefficients represent the posterior probability distribution.
Parameters of the posterior distribution are identified that quantify credibility of the simulator model.
The proposed method is demonstrated for calibration of a simulator of quasi-one-dimensional flow through a divergent nozzle. \bigskip


\begin{classcode} 15A29, 33C45, 58C40, 60G15, 62F15 \end{classcode} \bigskip

\centerline{\bfseries Nomenclature}\medskip
\begin{multicols}{2}
%\hbox to \textwidth{\hsize\textwidth\vbox{\hsize18pc
\hspace*{-12pt} {} {\bfseries Symbols} \\
\begin{tabular}{ll}
\hspace*{7pt} {A} & Nozzle cross sectional area\\
\hspace*{7pt} {C} & Covariance function\\
\hspace*{7pt} {d,l,s,c} & Spectral expansion coefficients \\
\hspace*{7pt} {$e(\cdot,\cdot)$} & Eigenfunctions\\
\hspace*{7pt} {E} & Total energy\\
\hspace*{7pt} {H} & Hermite polynomial \\
\hspace*{7pt} {L} & No. of Legendre polynomials\\
\hspace*{7pt} {M} & No. of system responses \\
\hspace*{7pt} {N} & No. of eigenfunctions \\
\hspace*{7pt} {P} & Static pressure \\
\hspace*{7pt} {$\mathcal{P}(\cdot)$} & Probability \\
\hspace*{7pt} {$T(\cdot,\cdot)$} & System model \\
\hspace*{7pt} {$u(\cdot,\cdot)$} & Subsystem model \\
\hspace*{7pt} {v} & Velocity \\
\hspace*{7pt} {w} & Weights of a Gaussian quadrature \\
\hspace*{7pt} {y} & System response \\
\hspace*{7pt} {$\mathbf{Y}$} & Set of system responses \\
\hspace*{7pt} {$\alpha, \beta$} & Hyper-parameters of Gamma/  \\
~ & Inverse Gamma distribution\\
\hspace*{7pt} {$\delta(\cdot)$} & Discrepancy function \\
\end{tabular} \\
%\hspace*{-12pt} {} {\bfseries Greek Symbols} \\
\begin{tabular}{ll}
\hspace*{7pt} {$\epsilon(\cdot)$} & Experimental uncertainty\\
\hspace*{7pt} {$\zeta(\cdot,\cdot)$} & True system response \\
%\end{tabular} \\
%\newpage
%\begin{tabular}{ll}
\hspace*{7pt} {$\boldsymbol{\theta}$} & Uncertain hyper-parameters \\
~ & of random function \\
\hspace*{7pt} {$\lambda$} & Eigenvalues \\
\hspace*{7pt} {$\boldsymbol{\mu}$} & Mean vector \\
\hspace*{7pt} {$\boldsymbol{\xi}$} & Vector of standard normal \\
~ & random variables \\
\hspace*{7pt} {$\rho$} & Density \\
\hspace*{7pt} {$\sigma$} & Standard deviation \\
\hspace*{7pt} {$\Sigma$} & Covariance matrix \\
\hspace*{7pt} {$\chi$} & Random variable in KL \\ 
~ & expansion \\
\hspace*{7pt} {$\phi, \psi$} & Scaled Legendre polynomial \\
\hspace*{7pt} {$\left\langle \cdot, \cdot \right\rangle$} & Inner product \\
\end{tabular} \\
\hspace*{-12pt} {} {\bfseries Subscript} \\
\begin{tabular}{ll}
\hspace*{7pt} {e} & $~~~~$ Experimental observation \\
\hspace*{7pt} {q} & $~~~~$ Quadrature node \\
\hspace*{7pt} {u} & $~~~~$ Subsystem\\
\hspace*{7pt} {$\delta$} & $~~~~$ Discrepancy function \\
\end{tabular} \\
\hspace*{-12pt} {} {\bfseries Superscript} \\
\begin{tabular}{ll}
\hspace*{7pt} {$\hat{\cdot}$} & $~~~~~$ Polynomial chaos coefficient \\
\end{tabular}
\end{multicols}
\end{abstract}

\section{Introduction}
\begin{figure}[tp]
  \begin{center}
   \includegraphics[width = .9\columnwidth] {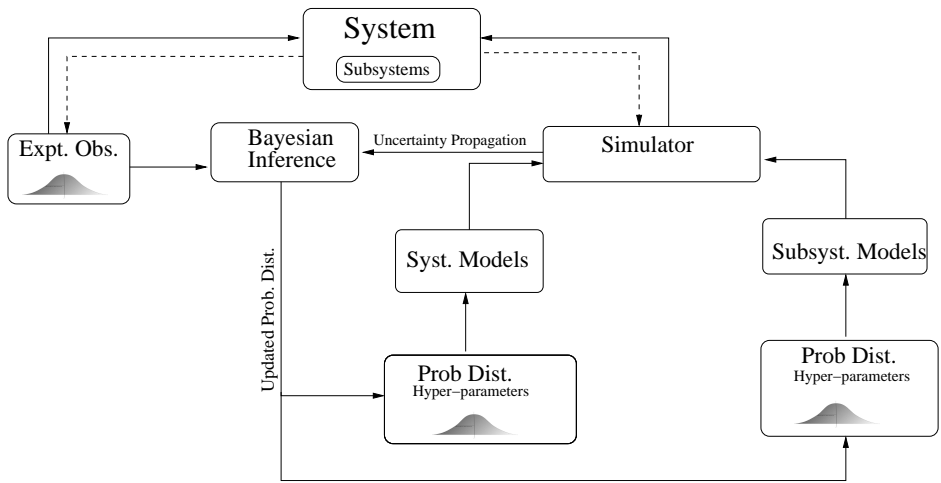}
  \end{center}
 \caption{Conceptual Architecture of Bayesian Calibration for Uncertain Models}
\label{FigSim}
\end{figure}

With the present ubiquitous use of computer simulators for the scientific investigations, uncertainty quantification and calibration of the simulator models is identified as an important area of research \cite{Mehta91,Mehta96,OreskesScience94,Mehta98,OberkampfRESS02,ThunPhD,cheung11}.
Significant developments of the last decade have established the Bayesian framework as a preferred method for uncertainty quantification and calibration of computer simulators \cite{TrucanoRESS06,Glim99,KennedyJRSS01,HigdonJSC04,Goldstein04,Bayarri_Tech07,higdon_jasa08,KellyRESS09,goldstein_jspi09}.
This paper explores a Bayesian framework for calibration and credibility assessment of a computer simulator.
The framework is particularly developed for simulators with uncertain subsystem models that are represented using functions.
Figure \ref{FigSim} shows schematic of the proposed framework for a complex system consisting of physically or mechanically interconnected subsystems.
The system is investigated using experimental observations and computer simulators, that use available information about the system for initial setup, while, repeated runs of the experiment and the simulator are used to understand more about the system.
Experimental observations of the system response are used in the Bayesian framework for calibration of the computer simulator.

Most appealing facet of the Bayesian framework is its ability to provide the complete posterior statistics.
However barring very simple cases, statistical sampling techniques are required for solution of the Bayesian calibration and uncertainty propagation problems.
Markov Chain Monte Carlo (MCMC) method \cite{BesagSS95,Gamerman} is one of the most widely used sampling technique for the Bayesian calibration.
However, exploration of the posterior distribution using the MCMC requires collection of a large number of samples for satisfactory approximation (often in the range of $10^3-10^6$), rendering the Bayesian framework computationally prohibitive for a large scale system simulator.
Thus, for generalized application of a Bayesian framework to the complex large scale system simulators, development of a computationally efficient Bayesian calibration technique is essential.

Marzouk et al. \cite{MarzoukJCP07} have proposed a spectral projection based method for computationally efficient Bayesian calibration.
The method uses a spectral expansion of a prior in generalized Polynomial Chaos basis.
The Polynomial Chaos based spectral projection method is extensively investigated in the literature as a computationally efficient alternative to statistical methods for uncertainty propagation with comparable accuracy \cite{Walters02}.
Polynomial chaos method is based on a concept of Homogeneous Chaos introduced by Wiener \cite{WienerAJM38,Wiener}, where a random variable is spectrally expanded in terms of Hermite polynomials.
Cameron and Martin \cite{CameronAM47} have shown that any non-linear functional can be expanded in terms of a series of Hermite polynomials in $L^2$ sense.
Although earlier attempts at using the polynomial chaos method (especially for turbulent fluid flow modeling) were not very successful \cite{MeechamJFM68,OrszagPF67,ChorinJFM74}, the method is found to be useful for solution of stochastic finite element \cite{Ghanem91,GhanemPhyD99,Ghanem} and stochastic fluid flow problems \cite{Knio_FDR_2006, Maitre_2001}.
Xiu and Karniadakis \cite{Xiu_SJSC_2004, XiuJCP03} have generalized the polynomial chaos method for spectral projection in terms of the Askey scheme of polynomials \cite{Koekoek}.
The generalized Polynomial Chaos (gPC) method have been applied by various researchers for uncertainty propagation through simulators of systems of engineering importance \cite{Lucor03,Mathelin04,Narayanan04,Najm_afm}.

The method proposed by Marzouk et al. \cite{MarzoukJCP07} uses the gPC for propagation of the prior uncertainty to the simulator predictions.
The resultant gPC expansion of the simulator predictions is used to define the likelihood.
On availability of the experimental observations, probability distribution of the gPC basis is updated using the Bayesian calibration.
On substitution of respective polynomial chaos coefficients, posterior distribution of parameters is obtained.
Marzouk et al. \cite{MarzoukJCP09} have further extended the method for inference of spatially/temporally varying uncertain parameters.

In this paper, the method proposed by Marzouk et al. \cite{MarzoukJCP09} is extended for the prior with uncertain hyper-parameters.
Though a family of probability distribution to represent the prior uncertainty can be specified, associated hyper-parameters are rarely known deterministically.
Realistic quantification of the prior uncertainty requires specification of probability distribution for uncertain hyper-parameters.
Use of the uncertain hyper-parameters is more ubiquitous in case of calibration of simulators with model structural uncertainty.
Hierarchical Bayesian inference is proposed in the literature for calibration in presence of uncertain hyper-parameters \cite{KennedyJRSS01,HigdonJSC04}.
However, methodology proposed by Marzouk et al. \cite{MarzoukJCP07,MarzoukJCP09} does not explicitly consider the effect of uncertain hyper-parameters in the formulation.
To make the spectral stochastic projection based Bayesian inference more precise, it is necessary to include uncertain hyper-parameters in the formulation.

This paper proposes an extension of the method of Marzouk et al. \cite{MarzoukJCP07,MarzoukJCP09} to take into consideration the uncertainty in hyper-parameters of the prior distribution.
A methodology is proposed to obtain Karhunnen-Loeve expansion (KL expansion) of a stochastic process in terms of functions of the hyper-parameters.
The prior uncertainty in hyper-parameters is expanded in the gPC basis.
Galerkin projection is used to evaluate gPC coefficients of the resultant KL expansion terms of a stochastic process.
The prior uncertainty in subsystem model, represented in terms of gPC basis, is propagated to the simulator predictions using the intrusive Galerkin projection approach \cite{Ghanem}.
The Bayesian calibration is reformulated as a MCMC sampling from the posterior distribution of the gPC basis.
The resultant gPC expansion with posterior distribution of the basis defines the posterior distribution of the uncertain parameters and the model structure.
The posterior distribution of the model structure defines credibility of the simulator model.
The posterior parameters are identified that quantifies acceptability of the simulator.
The proposed method is demonstrated using a simulator of a quasi-one-dimensional flow through a nozzle.
The particular choice of the application is motivated by the fact that the quasi-one-dimensional nozzle flow is well understood and can be simulated with limited computational resources.

This research  extends existing state of the art by: (a) %\begin{itemize} 
%\item 
extending the gPC expansion based method of Marzouk et al. \citet{MarzoukJCP07,MarzoukJCP09} for priors with uncertain hyper-parameters; and (b) %\item 
providing guidelines for acceptability of the simulator model using hyper-parameters of the posterior distribution. 
Note that a preliminary version of this work was reported in \cite{TagadeIDETC11}, while this article is significantly expanded by including (a) model structural uncertainty; (b) substantially elaborate theoretical analysis; and (c) additional numerical results to establish computational efficiency and efficacy of the method.

The rest of the paper is organized as follows. Section 2 provides statistical formulation of the problem.
In section 3, proposed method is discussed in detail.
In section 4, numerical results for the calibration of a quasi-one-dimensional nozzle flow simulator are presented and finally in section 5, the paper is summarized and concluded.

\section{Statistical Formulation}

Let the system be investigated by observing $M$ system responses, while, $T_j({\mathbf x},u({\mathbf x}_s))$ be an available simulator of the $j^{th}$ system response, where ${\mathbf x} \in \mathcal{X}$ is a set of deterministic control inputs and $u({\mathbf x}_s)$ are uncertain subsystem model.
Note that $u({\mathbf x}_s)$ are functions with argument ${\mathbf x}_s$ consisting of mix of some of the elements of ${\mathbf x}$, some intermediate calculations from the system model and independent subsystem specific input parameters.
Output of the subsystem model is used in the system model for further calculations. 
For brevity, the discussion presented in this paper assumes a single uncertain subsystem model; however, the proposed method can be extended to a more generic case without any change.  
Let $u_t({\mathbf x}_s)$ denote the `true' subsystem model, while, $\zeta_j({\mathbf x},u_t({\mathbf x}_s))$ be the `true' but unknown $j^{th}$ system response prediction.
Note that the simulator $T_j(\cdot,\cdot)$ approximates the system response within limits of available knowledge. 
Thus conditional on the true subsystem model $u_t({\mathbf x}_s)$, $T_j({\mathbf x},u_t({\mathbf x}_s))$ deviates from $\zeta_j({\mathbf x},u_t({\mathbf x}_s))$ by 
\begin{equation}
\zeta_j({\mathbf x},u_t({\mathbf x}_s)) = T_j({\mathbf x},u_t({\mathbf x}_s)) + \delta_j({\mathbf x}),
\label{zeta1}
\end{equation}
where $\delta_j({\mathbf x})$ is a discrepancy function.  

Let $y_{e_j}(\mathbf{x})$ represent an experimental observation of the system at a control input setting $\mathbf{x}$, while $\epsilon_j(\mathbf{x})$ denote the corresponding measurement uncertainty.
Relationship between experimental observation and the true system response is given by
\begin{equation}
y_{e_j}(\mathbf{x}) = \zeta_j(\mathbf{x},u_t(\mathbf{x}_s)) + \epsilon_j(\mathbf{x}).
\label{ExpTrue}
\end{equation}
Let experimental observations are obtained at $N$ input conditions and denote a set of experimental observations by $\mathbf{Y}_e = \{\mathbf{Y}_{e_j}; ~ j=1,...,M \}$ where $\mathbf{Y}_{e_j} = \{y_{e_j}(\mathbf{x}_i); ~ i=1,...,N \}$.
Similarly define $\boldsymbol{\delta}_j=\{\delta_j(\mathbf{x}_i); ~ i=1,...,N \}$ and $\boldsymbol{\delta}=\{\boldsymbol{\delta}_j; ~ j=1,...,M\}$.
Also define $\mathbf{u}_t = \{u_t(\mathbf{x}_{s,i});~i=1,...,N_u\}$, where typically $N_u$ is significantly greater than $N$. 
Note that $\mathbf{u}_t$ represent a realization of a random function $u_t(\mathbf{x}_{s})$ at $N_u$ input settings. 
Information available from the experimental observations is used in the Bayes theorem as
\begin{equation}
\mathcal{P}(\mathbf{u}_t,\boldsymbol{\delta}\mid\mathbf{Y}_e) \propto \mathcal{P}(\mathbf{Y}_e\mid \mathbf{u}_t,\boldsymbol{\delta}) \times \mathcal{P}(\mathbf{u}_t,\boldsymbol{\delta}),     
\label{postpredprob}
\end{equation}
where $\mathcal{P}(\mathbf{u}_t,\boldsymbol{\delta})$ is a prior, $\mathcal{P}(\mathbf{Y}_e\mid \mathbf{u}_t, \boldsymbol{\delta})$ is likelihood and $\mathcal{P}(\mathbf{u}_t,\boldsymbol{\delta} \mid\mathbf{Y}_e)$ is a posterior probability distribution.

Prior probability in $\mathbf{u}_t$ and $\boldsymbol{\delta}$ is specified using independent Gaussian processes.   
However, complete definition of the prior requires specification of uncertain hyper-parameters of the probability distribution.
Let $\boldsymbol{\theta}_u \in \mathbf{\Theta}_u$ and $\boldsymbol{\theta}_\delta \in \mathbf{\Theta}_\delta$ are the hyper-parameters of subsystem model and discrepancy function respectively, where $\mathbf{\Theta}_u$ and $\mathbf{\Theta}_\delta$ are set of possible values.
Thus in the presence of the uncertain hyper-parameters, posterior probability distribution (\ref{postpredprob}) takes the form
\begin{equation}
\begin{split}
& \mathcal{P}(\mathbf{u}_t(\boldsymbol{\theta}_u),\boldsymbol{\delta}(\boldsymbol{\theta}_\delta),\boldsymbol{\theta}_u,\boldsymbol{\theta}_\delta \mid \mathbf{Y}_e) \propto  \mathcal{P}(\mathbf{Y}_e \mid \mathbf{u}_t(\boldsymbol{\theta}_u),\boldsymbol{\delta}(\boldsymbol{\theta}_\delta),\boldsymbol{\theta}_u,\boldsymbol{\theta}_\delta) \\
& \qquad \times \mathcal{P}(\mathbf{u}_t(\boldsymbol{\theta}_u), \boldsymbol{\delta}(\boldsymbol{\theta}_\delta) \mid \boldsymbol{\theta}_u, \boldsymbol{\theta}_\delta) \times \mathcal{P}(\boldsymbol{\theta}_u) \times \mathcal{P}(\boldsymbol{\theta}_\delta) .
\end{split}
\label{postpredprob_hp}
\end{equation}
For better readability, dependence of the uncertain subsystem model and the discrepancy function on uncertain hyper-parameters is explicitly shown henceforth.

In the present paper, Bayesian inference is developed assuming the Gaussian process prior for discrepancy functions, as it is extensively used in the literature for specification of prior on random functions \cite{SacksSS89,KennedyJRSS01,PauloAS05,O'HaganRESS06}. 
Prior uncertainty in subsystem model and discrepancy function is assumed to be independent.     
Uncertainty in the experimental observations is specified by a zero mean normally distributed random variable $\epsilon_j$ with standard deviation $\sigma_{e_j}$.
The experimental uncertainty at different input settings $\mathbf{x}$ is assumed to be uncorrelated.
On marginalization of $\delta_j(\mathbf{x})$, posterior distribution is given by
\begin{equation}
\begin{split}
& \mathcal{P}(\mathbf{u}_t(\boldsymbol{\theta}_u),\boldsymbol{\delta}(\boldsymbol{\theta}_u),\boldsymbol{\theta}_u,\boldsymbol{\theta}_\delta \mid \mathbf{Y}_e) \propto \prod^{M}_{j=1} \frac{1}{\sqrt{\mid \Sigma_j \mid}} exp \left\{ -\frac{1}{2} \left(\mathbf{Y}_{e_j} - \boldsymbol{\mu}_j \right)^T \Sigma^{-1}_j \left(\mathbf{Y}_{e_j} - \boldsymbol{\mu}_j \right) \right\}  \\
& \qquad \times ~ \mathcal{P}(\mathbf{u}_t(\boldsymbol{\theta}_u) \mid \boldsymbol{\theta}_u) \times \mathcal{P}(\boldsymbol{\theta}_u) \times \mathcal{P}(\boldsymbol{\theta}_\delta),     
\end{split}
\label{bayes_fin}
\end{equation}
where $\boldsymbol{\mu}_j = \{T_j(\mathbf{x}_i,u_t(\mathbf{x}_s;\boldsymbol{\theta}_u)) + E(\delta_j(\mathbf{x}_i;\boldsymbol{\theta}_\delta)); ~ i=1,...,N\}$ while $\Sigma_j = \Sigma_{\delta_j} + \sigma^2_j I_N$, $I_N$ being $N\times N$ identity matrix.

Metropolis-Hastings algorithm \cite{MetropolisJCP53,HastingsBio70} is used to sample from the posterior distribution (\ref{bayes_fin}).
For each sample, evaluation of $T_j(\mathbf{x},u_t(\mathbf{x}_s;\boldsymbol{\theta}_u))$, $\Sigma^{-1}_j$ and $\mid \Sigma_j \mid$ impose significant computational expenses on the Bayesian framework.
In the present paper, a gPC expansion based method is proposed for computationally efficient evaluations of $T_j(\mathbf{x},u_t(\mathbf{x}_s;\boldsymbol{\theta}_u))$, $\Sigma^{-1}_j$ and $\mid\Sigma_j\mid$.

\section{Generalized Polynomial Chaos Expansion of a Gaussian Process with Uncertain Hyper-parameters}
For brevity, the proposed spectral formulation is described first for a susbsystem model, $u(\mathbf{x};\boldsymbol{\theta}_u)$, which is subsequently extended for a discrepancy function.\footnote{For notational convenience, $\mathbf{x}_s$ is replaced by $\mathbf{x}$ in this section.}    
The formulation is derived for a zero mean Gaussian process, though, it can easily be extended for non-zero mean processes.
The derivation uses KL expansion of the Gaussian process with uncertain hyper-parameters, which is projected on a gPC basis using the intrusive Galerkin projection.   

\subsection{KL Expansion}
Let $u(\mathbf{x};\boldsymbol{\theta}_u)$ be a zero-mean Gaussian process with a covariance function $C_u(\mathbf{x}_1,\mathbf{x}_2;\boldsymbol{\theta}_u)$.
The covariance function can be approximated as \cite{Ghanem}
\begin{equation}
C_u(\mathbf{x}_1,\mathbf{x}_2;\boldsymbol{\theta_u}) = \sum^{N}_{n=1} \lambda_n(\boldsymbol{\theta_u}) e_n(\mathbf{x}_1,\boldsymbol{\theta_u}) e_n(\mathbf{x}_2,\boldsymbol{\theta_u}),
\label{MercTh}
\end{equation}
where $N$ is the number of expansion terms retained in the spectral approximation. 
$\lambda_n(\boldsymbol{\theta}_u)$ and $e_n(\mathbf{x},\boldsymbol{\theta}_u)$ are eigenvalues and eigenfunctions of the covariance kernel, which are given by solution of the Fredholm's integral equation of the second kind \cite{fie_09}
\begin{equation}
\int_{\mathcal{X}} C_u(\mathbf{x}_1,\mathbf{x}_2;\boldsymbol{\theta}_u) e_n(\mathbf{x}_1,\boldsymbol{\theta}_u) d \mathbf{x}_1 = \lambda_n(\boldsymbol{\theta}_u) e_n(\mathbf{x}_2,\boldsymbol{\theta}_u).
\label{fie}
\end{equation}
Explicit dependence of the eigenvalues and eigenfunctions on the hyper-parameters $\boldsymbol{\theta}_u$ should be noted.
The resultant KL expansion using (\ref{MercTh}) is given by \cite{Ghanem}
\begin{equation}
u(\mathbf{x};\boldsymbol{\theta}_u) = \sum^{N}_{n=1} \sqrt{\lambda_n(\boldsymbol{\theta}_u)} e_n(\mathbf{x},\boldsymbol{\theta}_u) \chi_n,
\label{kl_expn}
\end{equation}
where $\chi_n$ are independent zero-mean standard normal random variables.
For a given $\boldsymbol{\theta}_u$, the eigenvalue problem (\ref{fie}) can be numerically solved using a Galerkin projection based approach \cite{Huang_ijnme01}. 
In this paper, the approach is extended for uncertain hyper-parameters as follows.  

Eigenfunctions $e_n(\mathbf{x},\boldsymbol{\theta}_u)$ can be spectrally approximated as
\begin{equation}
e_n(\mathbf{x},\boldsymbol{\theta}_u) = \sum^{N}_{i=1} d^n_i(\boldsymbol{\theta}_u) \psi_i(\mathbf{x}),
\label{efun_approx}
\end{equation}
where $\psi_i(\mathbf{x})$ are Legendre polynomials and $d^n_i(\boldsymbol{\theta}_u)$ are respective expansion coefficients.
Use (\ref{efun_approx}) in (\ref{fie}), multiply both sides by $\psi_j(\mathbf{x}_2)$ and integrate w.r.t. $d \mathbf{x}_2$ to obtain
\begin{equation}
\begin{split}
&\sum^{N}_{i=1} d^n_i(\boldsymbol{\theta}_u) \int_{\mathcal{X}} \int_{\mathcal{X}} C_u(\mathbf{x}_1,\mathbf{x}_2;\boldsymbol{\theta}_u) \psi_i(\mathbf{x}_1) \psi_j(\mathbf{x}_2) d\mathbf{x}_1 d\mathbf{x}_2 \\
&\qquad =  \lambda_n(\boldsymbol{\theta}_u) \sum^{N}_{i=1} d^n_i(\boldsymbol{\theta}_u) \int_{\mathcal{X}} \psi_i(\mathbf{x}_2) \psi_j(\mathbf{x}_2) d\mathbf{x}_2.
\end{split}
\label{fie2}
\end{equation}
Using
\begin{align*}
&A_{ij}(\boldsymbol{\theta}_u)   =  \int_{\mathcal{X}} \int_{\mathcal{X}} C(\mathbf{x}_1,\mathbf{x}_2;\boldsymbol{\theta}_u) \psi_i(\mathbf{x}_1) \psi_j(\mathbf{x}_2) d\mathbf{x}_1 d\mathbf{x}_2,  & D_{ij}(\boldsymbol{\theta}_u) = d^j_i(\boldsymbol{\theta}_u), \\
&B_{ij}(\boldsymbol{\theta}_u)  = \int_{\mathcal{X}} \psi_i(\mathbf{x}_2) \psi_j(\mathbf{x}_2) d\mathbf{x}_2, &
\Lambda_{ii}(\boldsymbol{\theta}_u) =  \lambda_i(\boldsymbol{\theta}_u),
\label{abdl}
\end{align*}
(\ref{fie2}) can be written in a matrix form as
\begin{equation}
A(\boldsymbol{\theta}_u) D(\boldsymbol{\theta}_u) = \Lambda(\boldsymbol{\theta}_u) B(\boldsymbol{\theta}_u) D(\boldsymbol{\theta}_u),
\label{gep}
\end{equation}
which is a generalized eigenvalue problem (GEP) that can be solved using the QZ algorithm \cite{Ghanem}.

The matrices in (\ref{gep}) are functions of $\boldsymbol{\theta}_u$. Thus, to solve (\ref{gep}), consider spectral expansion of eigenvalues and eigenfunctions as
\begin{equation}
\lambda_n(\boldsymbol{\theta}_u) = \sum^{L}_{i=1} l^n_i \phi_i(\boldsymbol{\theta}_u), \qquad d^n_{k}(\boldsymbol{\theta}_u) = \sum^{L}_{i=1} c^{n}_{i,k} \phi_i(\boldsymbol{\theta}_u),
\label{hp_exp}
\end{equation}
where $\phi_i(\boldsymbol{\theta}_u)$ are appropriately scaled Legendre polynomials that form complete orthonormal basis on $L^2(\boldsymbol{\Theta}_u)$.
The coefficients are given by
\begin{equation}
l^n_i = \frac{\int_{\boldsymbol{\Theta}_u} \lambda_n(\boldsymbol{\theta}_u) \phi_i(\boldsymbol{\theta}_u) d\boldsymbol{\theta}_u}{\int_{\boldsymbol{\Theta}_u} \phi^2_i(\boldsymbol{\theta}_u) d\boldsymbol{\theta}_u} , \qquad c^n_{i,k} = \frac{\int_{\boldsymbol{\Theta}_u} d^n_{k}(\boldsymbol{\theta}_u) \phi_i(\boldsymbol{\theta}_u) d\boldsymbol{\theta}_u}{\int_{\boldsymbol{\Theta}_u} \phi^2_i(\boldsymbol{\theta}_u) d\boldsymbol{\theta}_u}.
\label{hp_exp_coef}
\end{equation}
Using the Gauss-Legendre quadrature, the integrals can be approximated as
\begin{equation}
\begin{split}
&\int_{\boldsymbol{\Theta}_u} \lambda_i(\boldsymbol{\theta}_u) \phi_i(\boldsymbol{\theta}_u) d\boldsymbol{\theta}_u = \sum^{N_q}_{q=1} \lambda_i(\boldsymbol{\theta}^q_u) \phi_i(\boldsymbol{\theta}^q_u) w_q, \\
&\int_{\boldsymbol{\Theta}_u} d^n_k(\boldsymbol{\theta}_u) \phi_i(\boldsymbol{\theta}_u) d\boldsymbol{\theta}_u = \sum^{N_q}_{q=1} d^n_{k}(\boldsymbol{\theta}^q_u) \phi_i(\boldsymbol{\theta}^q_u) w_q
\end{split}
\label{GaussQuad_hp}
\end{equation}
where $N_q$ are the number of quadrature points used, $\boldsymbol{\theta}^q_u$ are the quadrature nodes while $w_q$ are the respective quadrature weights.
The expansion coefficients, $l^n_i$ and $c^n_{i,k}$, are calculated by solving (\ref{gep}) at quadrature nodes $\boldsymbol{\theta}^q_u$ and substituting the solution in (\ref{GaussQuad_hp}) to evaluate integrals in (\ref{hp_exp_coef}).       
%----------------------------------------------------------------------------  
\subsection{gPC Expansion of a Gaussian Process}                          
%----------------------------------------------------------------------------  
Hyper-parameters $\boldsymbol{\theta}_u$ can be expanded in gPC basis as \cite{XiuJCP03}
\begin{equation}
\boldsymbol{\theta}_u = \sum^{P}_{p=1} \hat{\theta}^u_p H_p(\boldsymbol{\xi}),
\end{equation}
where $H_p(\boldsymbol{\xi})$ are the Hermite polynomials, $\hat{\theta}^u_p$ are respective expansion coefficients, while $\boldsymbol{\xi}$ is a vector of independent standard normal random variables.
Using the intrusive approach, gPC expansion of $\phi_i(\boldsymbol{\theta}_u)$ is given by
\begin{equation}
\phi_i(\boldsymbol{\theta}_u) = \sum^{P}_{p=1} \hat{\phi}^i_p H_p(\boldsymbol{\xi}).
\label{pc_phi}
\end{equation}

Using (\ref{hp_exp}) and (\ref{pc_phi}), gPC expansion of $\sqrt{\lambda_n(\boldsymbol{\theta}_u)}$ is given by
\begin{equation}
\sqrt{\lambda_n(\boldsymbol{\theta}_u)} = \sum^{L}_{i=1} \sum^{P}_{p=1}  s^n_i \hat{\phi}^i_p H_p(\boldsymbol{\xi}),   
\label{gpc_lmbd}
\end{equation}
where
$
s^n_k = \big\langle \sqrt{\sum^{L}_{i=1} l^n_i \phi_i(\boldsymbol{\theta}_u)}, \phi_k(\boldsymbol{\theta}_u) \big\rangle / \left\langle \phi^2_k(\boldsymbol{\theta}_u) \right\rangle,
$
and $\left\langle \cdot, \cdot \right\rangle$ denote an inner product\footnote[2]{For example, inner product of functions $f(\cdot)$ and $g(\cdot)$ is given by 
\begin{equation*}
\left\langle f(\cdot), g(\cdot) \right\rangle = \int_{\cdot \in \mathcal{O}} f(\cdot) g(\cdot) d \mu(\cdot), 
\end{equation*}
where $\mu(\cdot)$ is a measure on $L^2(\mathcal{O})$}.
Similarly, the gPC expansion of eigenfunctions can be obtained using (\ref{hp_exp}) and (\ref{pc_phi}) as
\begin{equation}
e_n(\mathbf{x},\boldsymbol{\theta}_u) = \sum^{N}_{k=1} \sum^L_{i=1}  \sum^{P}_{p=1} c^{n}_{i,k} \hat{\phi}^i_p H_p(\boldsymbol{\xi}) \psi_k(\mathbf{x}).
\label{gpc_en}
\end{equation}
Thus, a zero-mean Gaussian process $u(\mathbf{x};\boldsymbol{\theta}_u)$ can be expanded in gPC basis as
\begin{equation}
u(\mathbf{x};\boldsymbol{\theta}_u) = \sum^P_{p=1} \hat{u}_p(\mathbf{x}) H_p(\boldsymbol{\xi}),
\label{gpc_gp}
\end{equation}
where the expansion coefficients $\hat{u}_k(\mathbf{x})$ are given by using (\ref{kl_expn}), (\ref{gpc_lmbd}) and (\ref{gpc_en}) as
\begin{equation}
\hat{u}_k(\mathbf{x}) = \sum^{N}_{n=1} \sum^{N}_{m=1} \sum^L_{i=1} \sum^L_{j=1} \sum^P_{p=1} \sum^P_{q=1} s^n_i c^n_{j,m} \hat{\phi}^i_p \hat{\phi}^j_q \psi_m(\mathbf{x}) \frac{\left\langle H_p(\boldsymbol{\xi}), H_q(\boldsymbol{\xi}), H_{n+1}(\boldsymbol{\xi}), H_k(\boldsymbol{\xi}) \right\rangle}{\left\langle H^2_k(\boldsymbol{\xi})\right\rangle}
.
\label{gpc_gp_coef}
\end{equation}
Note that since $u(\mathbf{x};\boldsymbol{\theta}_u)$ is a Gaussian process, $\chi_n$ in (\ref{kl_expn}) are standard normal variables,
thus,
$\chi_n = H_{n+1}(\boldsymbol{\xi})$
is used in (\ref{gpc_gp_coef}).

\section{Stochastic Spectral Projection based Bayesian Calibration}
The prior uncertainty in the subsystem model, $u_t(\mathbf{x}_s;\boldsymbol{\theta}_u)$, and the discrepancy function, $\delta_j(\mathbf{x};\boldsymbol{\theta}_\delta)$, is given by independent Gaussian processes.
Using (\ref{gpc_gp}), spectral expansion of the prior is given by
\begin{equation}
u_t(\mathbf{x}_s;\boldsymbol{\theta}_u) = \sum^{P}_{p=1} \hat{u}_p(\mathbf{x}_s) H_p(\boldsymbol{\xi});\qquad
\delta_j(\mathbf{x};\boldsymbol{\theta}_\delta) = \sum^{P}_{p=1} \hat{\delta}^j_p(\mathbf{x}) H_p(\boldsymbol{\xi}).
\label{gPC_u_delta}
\end{equation}
Intrusive Galerkin projection approach is used to propagate the prior uncertainty in $u_t(\mathbf{x}_s;\boldsymbol{\theta}_u)$ to the simulator predictions, thus, 
\begin{equation}
T_j(\mathbf{x},u_t(\mathbf{x}_s;\boldsymbol{\theta}_u)) = \sum^{P}_{p=1} \hat{T}^j_p(\mathbf{x}) H_p(\boldsymbol{\xi}).   
\end{equation}
Using the gPC expansion of $T_j(\mathbf{x},u_t(\mathbf{x}_s;\boldsymbol{\theta}_u))$ and $\delta_j(\mathbf{x};\boldsymbol{\theta}_\delta)$ in (\ref{zeta1}) to obtain the gPC expansion of $\zeta_j(\mathbf{x},u_t(\mathbf{x}_s;\boldsymbol{\theta}_u))$ as
\begin{equation}
\zeta_j(\mathbf{x},u_t(\mathbf{x}_s;\boldsymbol{\theta}_u))=\sum^{P}_{p=1} \hat{\zeta}^j_p (\mathbf{x}) H_p(\boldsymbol{\xi}),
\label{pc_zeta}
\end{equation}
where
%\begin{equation}
$\hat{\zeta}^j_p (\mathbf{x}) = \hat{T}^j_p(\mathbf{x}) + \hat{\delta}^j_p(\mathbf{x}).$
%\label{pc_zeta}
%\end{equation}
Use the gPC expansion (\ref{pc_zeta}) in (\ref{ExpTrue}) to obtain
\begin{equation}
y_{e_j}(\mathbf{x}) = \sum^{P}_{p=1} \hat{\zeta}^j_p(\mathbf{x}) H_p(\boldsymbol{\xi}) + \epsilon_j.
\label{Final_stat_form}
\end{equation}
Note that $\boldsymbol{\xi}$ are the only uncertain variables in (\ref{Final_stat_form}), thus, the Bayesian inference problem is reformulated as sampling from the posterior distribution of $\boldsymbol{\xi}$.

Let $\hat{\mathbf{T}}=\{\hat{T}^j_p(\mathbf{x}); ~ p=1,...,P; ~ j=1,...,M \}$ and $\hat{\boldsymbol{\delta}}=\{\hat{\delta}^j_p(\mathbf{x}); ~ p=1,...,P; ~ j=1,...,M \}$ define the sets of respective gPC coefficients.
Using the gPC expansion of the uncertain variables ((\ref{gPC_u_delta}) -- (\ref{Final_stat_form})) in (\ref{bayes_fin}), the Bayesian calibration problem can be reformulated in terms of $\boldsymbol{\xi}$ that capture all the randomness in the system model and hyper-parameters:
\begin{eqnarray}
\mathcal{P}(\boldsymbol{\xi} \mid \mathbf{Y}_e,\hat{\mathbf{T}},\hat{\boldsymbol{\delta}}) & \propto &  \mathcal{P}(\mathbf{Y}_e \mid \boldsymbol{\xi},\hat{\mathbf{T}},\hat{\boldsymbol{\delta}})  \times  \mathcal{P}(\boldsymbol{\xi}).
\label{bayes_fin_xi}
\end{eqnarray}
Since $\boldsymbol{\xi}$ are i.i.d. standard normal random variables, with independent Gaussian uncertainties in experimental data, the posterior distribution of $\boldsymbol{\xi}$ can be obtained by:
\begin{eqnarray}
\mathcal{P}(\boldsymbol{\xi} \mid \mathbf{Y}_e,\hat{\mathbf{T}},\hat{\boldsymbol{\delta}}) & \propto &\prod^{M}_{j=1} \frac{1}{\sqrt{| \Sigma_j |}} \exp \left\{ -\textstyle{\frac{1}{2}} \left(\mathbf{Y}_{e_j} - \mathbf{\mu}_j \right)^T \Sigma^{-1}_j \left(\mathbf{Y}_{e_j} - \mathbf{\mu}_j \right) \right\} \times  \prod^{N_d}_{n=1} e^{- \xi^2_n /2 },\qquad
\label{bay_final_form}
\end{eqnarray}
where $\boldsymbol{\mu}_j = \{\sum^{P}_{p=1} \hat{T}^j_p(\mathbf{x}_i) H_p(\boldsymbol{\xi}) + \hat{\delta}^j_0(\mathbf{x}_i); ~ i=1, \cdots ,N\}$.

Markov Chain Monte Carlo (MCMC) method is used to sample from (\ref{bay_final_form}).
Note that MCMC applied to (\ref{bay_final_form}) does not require solution of the simulation model, $T_j(\mathbf{x},u(\mathbf{x}_s,\boldsymbol{\theta}_u))$, thus, the posterior distribution can be explored efficiently.
However, solution of (\ref{bay_final_form}) requires  numerical evaluation of $| \Sigma_j |$ and $\Sigma^{-1}_j$, which may impose non-trivial computational cost on the MCMC sampling.
In the present paper, numerical evaluation of $| \Sigma_j |$ and $\Sigma^{-1}_j$ is accelerated as follows.

The numerical evaluation of $| \Sigma_j |$ and $\Sigma^{-1}_j$ is accelerated using the gPC expansion of the individual elements of the covariance matrix $\Sigma_{j}(\mathbf{x}_1,\mathbf{x}_2)$ as
\begin{equation}
\Sigma_{j}(\mathbf{x}_1,\mathbf{x}_2) = \sum^{P}_{p=1} \hat{C}^j_p(\mathbf{x}_1,\mathbf{x}_2) H_p(\boldsymbol{\xi}).
\label{pc_cov_mat}
\end{equation}
The determinant $| \Sigma_j |$ and the individual elements of the inverse $\Sigma^{-1}_j$ can also be expanded in gPC basis as
\begin{eqnarray}
| \Sigma_j |= \sum^{P}_{p=1} \hat{D}^j_p H_p(\boldsymbol{\xi}), \qquad \Sigma^{-1}_j(\mathbf{x}_1,\mathbf{x}_2) = \sum^{P}_{p=1} \hat{I}^j_p(\mathbf{x}_1,\mathbf{x}_2) H_p(\boldsymbol{\xi}),
\label{gpc_mat}
\end{eqnarray}
where the gPC expansion coefficients are given by
\begin{eqnarray}
\hat{I}^j_k =\frac{\langle \Sigma^{-1}_j, H_k(\boldsymbol{\xi})\rangle}{\langle H^2_k \rangle}, \qquad \hat{D}^j_k =\frac{\langle | \Sigma_j |, H_k(\boldsymbol{\xi})\rangle}{\langle H^2_k \rangle}.
\label{sgm_pc}
\end{eqnarray}
Gaussian quadrature  is used to evaluate polynomial chaos coefficients.
$\langle \Sigma^{-1}_{j}, H_k(\boldsymbol{\xi})\rangle$ and $\langle | \Sigma_{j}|, H_k(\boldsymbol{\xi})\rangle$ are calculated by evaluating $\Sigma^{-1}_j$ and $| \Sigma_j |$ at quadrature nodes, while $\hat{I}^j_k$ and $\hat{D}^j_k $ are evaluated using (\ref{sgm_pc}).
The resultant gPC expansion (\ref{gpc_mat}) is used in (\ref{bay_final_form}) for MCMC sampling.

%--------------------------------------------------------------------------------------------------------------------------------------
\section{Numerical Example: Calibration of Quasi-One-Dimensional Nozzle Flow Simulator}
\subsection{Problem Setup}
A quasi-one-dimensional supersonic flow through a nozzle is considered to verify the proposed Bayesian calibration method.
The quasi-one-dimensional flow is uniform across the cross-section with properties varying in $x$-direction, while, the effect of change in the cross sectional area is considered.
The flow is defined using the compressible Euler equations in conservative form
\begin{equation}
\frac{\partial \mathbf{q}}{\partial t} + \frac{\partial \mathbf{f}}{\partial x} = \mathbf{g},
\label{EulerEq}
\end{equation}
where,
\begin{eqnarray*}
\begin{tabular} {lll}
$ \mathbf{q} = \left(
   \begin{array}{c}
		\rho A\\
                \rho v A\\
                \rho E A
   \end{array}
 \right), $ &
$ \mathbf{f} = \left(
   \begin{array}{c}
		\rho v A\\
                \rho v^2 A + P A\\
                \rho v E A + P v A
   \end{array}
 \right),$
 &
$ \mathbf{g} = \left(
   \begin{array}{c}
		0\\
                P \frac{\partial A}{\partial x} \\
                0
   \end{array}
 \right).$
\\
\end{tabular}
\label{one_d_nozz}
\end{eqnarray*}
Here, $\rho$ is density, $v$ is velocity, $A$ is cross sectional area, $P$ is static pressure and $E$ is the total energy per unit mass.
In the governing equations, static pressure is substituted by the total energy per unit mass using
\begin{equation}
E = \frac{P}{(\gamma-1) \rho} + \frac{1}{2} v^2,
\end{equation}
while the ideal gas equation is used for the closure.

Variation of nozzle area $A$ is assumed to be uncertain, which is inferred using the Bayesian framework. Stationary Gaussian process with known mean profile is used as a prior for uncertain nozzle area. 
The procedure of obtaining the stochastic spectral formulation is summarized as follows.
Define $q_1 = \rho A$, $q_2 = \rho v A$, $q_3 = \rho E A$ and the corresponding gPC expansions
$$
q_1 = \sum^{P}_{p=1} \hat{q}_{1,p} H_p(\boldsymbol{\xi}); \qquad q_2 = \sum^{P}_{p=1} \hat{q}_{2,p} H_p(\boldsymbol{\xi}); \qquad q_3 = \sum^{P}_{p=1} \hat{q}_{3,p} H_p(\boldsymbol{\xi}).     
$$
Multiplying both the sides by $H_k(\boldsymbol{\xi})$ and taking the inner product, governing equations of a stochastic quasi-one-dimensional nozzle flow is given by
\begin{equation}
\frac{\partial \mathbf{Q}}{\partial t} + \frac{\partial \mathbf{F}}{\partial x} = \mathbf{G},
\label{EulerEqSt}
\end{equation}
where
\begin{eqnarray*}
\left(\mathbf{Q}\right)_p = \left(
   \begin{array}{c}
		\hat{q}_{1,p}\\
                \hat{q}_{2,p}\\
                \hat{q}_{3,p}
   \end{array}
 \right),
\end{eqnarray*}
\begin{eqnarray*}
\left(\mathbf{F}\right)_p = \left(
   \begin{array}{c}
		\hat{q}_{2,p} \\
                ~ \\
                \sum^{P}_{i=1} \sum^{P}_{j=1} \sum^{P}_{k=1} \hat{q}_{2,i} \hat{q}_{2,j} \hat{r}_k \frac{\langle H_i H_j H_k H_p\rangle}{\langle H^2_p\rangle} + \\
                \frac{\gamma-1}{\gamma}\left[ \hat{q}_{3,p} - \sum^{P}_{i=1} \sum^{P}_{j=1} \sum^{P}_{k=1} \hat{q}_{2,i} \hat{q}_{2,j} \hat{r}_k \frac{\langle H_i H_j H_k H_p\rangle}{\langle H^2_p\rangle} \right] \\
                ~ \\
                \gamma \sum^{N_p}_{i=1} \sum^{N_p}_{j=1} \sum^{N_p}_{k=1} \hat{q}_{2,i} \hat{q}_{3,j} \hat{r}_k \frac{\langle H_i H_j H_k H_p\rangle}{\langle H^2_p\rangle} - \\
                 \frac{\gamma (\gamma-1) }{2} \left[\sum^{P}_{i=1} \sum^{P}_{j=1} \sum^{P}_{k=1} \sum^{P}_{l=1} \sum^{P}_{m=1} \hat{q}_{2,i} \hat{q}_{2,j} \hat{q}_{2,k} \hat{r}_l \hat{r}_m \frac{\langle H_i H_j H_k H_l H_m H_p\rangle}{\langle H^2_p\rangle}  \right]
   \end{array}
 \right),
\end{eqnarray*}

\begin{eqnarray*}
\left( \mathbf{G} \right)_p = \left(
   \begin{array}{c}
		0\\
                ~ \\
                \frac{\gamma-1}{\gamma} \sum^{P}_{i=1} \sum^{P}_{j=1} \hat{q}_{3,i}{\partial A_j}{\partial x} \frac{\langle H_i H_j H_k H_p\rangle}{\langle H^2_p\rangle} - \\
               \frac{\gamma(\gamma-1)}{2 \gamma} \sum^{P}_{i=1} \sum^{P}_{j=1}  \sum^{P}_{k=1} \sum^{P}_{l=1} \hat{q}_{2,i} \hat{q}_{2,j} \hat{r}_k {\partial A_l}{\partial x} \frac{\langle H_i H_j H_k H_l H_p\rangle}{\langle H^2_p\rangle} \\
                ~ \\
                0
   \end{array}
 \right)
\end{eqnarray*}
and $r=\frac{1}{q_1}$. Note that the governing equations is a set of $3\times P$ partial differential equations.
These governing equations are numerically solved using the central difference scheme in the spatial dimension and the fourth order Runge-Kutta method in the temporal dimension.
A uniform grid with $\Delta x = 0.01$ is used in the spatial dimension, while, the time step $\Delta t = 0.0001$ is used for time integration.
The inner products involved in the governing equations are evaluated \emph{a-priory} using the Gauss-Hermite quadrature. 

\begin{figure}[t]
  \begin{center}
  \begin {tabular}{l l}
  \subfigure []{\includegraphics[width=2.5in, height=2.45in] {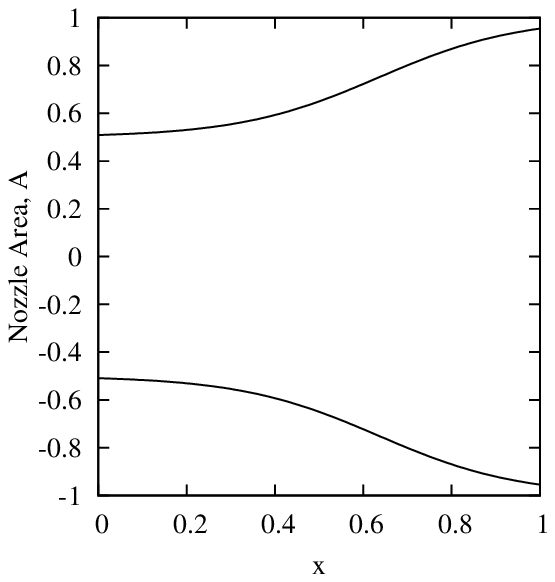}} &
    \subfigure [] {\includegraphics[width=2.5in, height=2.45in] {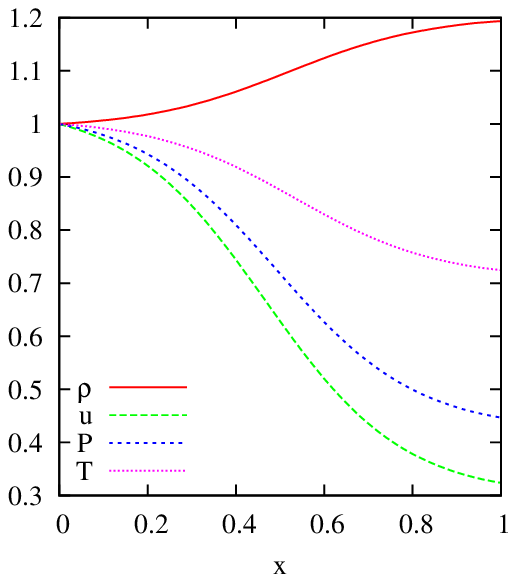}} \\
   \end{tabular}
  \end{center}
 \caption{Figure shows (a) Nozzle area variation and (b) steady state response predictions for deterministic case. All the steady state responses are normalized using inflow values.}
\label{noz_exp}
\end{figure}

For demonstration purpose, the proposed method is applied using `hypothetical test bed nozzle' data.
The `hypothetical nozzle' can be created using the nozzle simulator with sensors that can measure the nozzle response with typical accuracies.
A hypothetical nozzle is specified through a set of parameters and `true' area profile, all of which are treated as precisely known.
In all the test cases presented in this section, density $\rho$, velocity $v$, pressure $P$ and static temperature $T$ are used as responses of interest.
Steady state predictions using known nozzle area profile are used as experimental observations.
The experimental uncertainty is specified by a zero-mean normally distributed random variable with standard deviation given by 1\% of the mean value.
For all the test cases, inflow conditions are Mach number $M_{in} = 1.5$, static pressure $P_{in} = 1.0$ and density $\rho = 1.0$.
Figure \ref{noz_exp} shows a typical nozzle area profile and spatial variation of normalized responses at steady state.
Note that supersonic flow through a divergent nozzle results in increased velocity, while the static pressure, temperature and density are decreased.
As can be observed from Figure \ref{noz_exp}, this behavior is captured well by the nozzle simulator.
%%%%%%%%%
%%%%%%%%%

\subsection{Prior Uncertainty Propagation}
\begin{figure}[t]
  \begin{center}
  \begin {tabular}{l l}
  \subfigure []{\includegraphics[width=2.5in, height=2.45in] {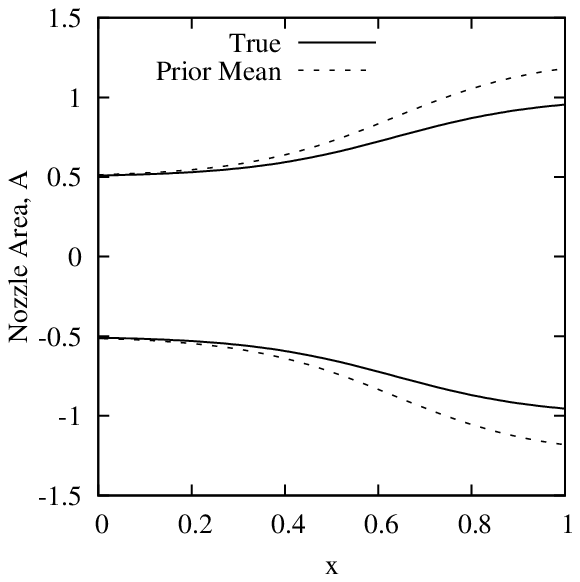}} &
    \subfigure [] {\includegraphics[width=2.5in, height=2.45in] {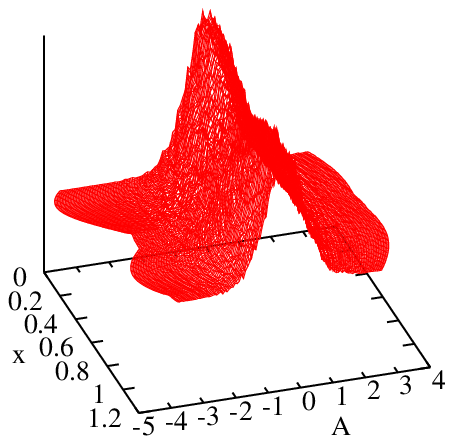}} \\
   \end{tabular}
  \end{center}
 \caption{Figure shows a) Comparison of prior mean nozzle area with true nozzle area and b) prior probability distribution of nozzle area.}
\label{noz_area_unc}
\end{figure}
%%%%%%%%%
Prior uncertainty in the nozzle area is specified using a Gaussian process with known mean and the covariance function is given by
\begin{equation}
C(x_1,x_2) = \sigma^2 \exp \left(-\lambda \left(x_1 - x_2\right)^2 \right),
\label{cov_noz_sq_exp}
\end{equation}
where $\sigma^2$ is the variance and $\lambda$ is the correlation length.
$\sigma^2$ and $\lambda$ are treated as uncertain.
Prior uncertainty in $\sigma^2$ is specified using an inverse Gamma distribution, $IG(9.0,0.5)$, while Gamma distribution, $G(5.0,0.2)$, is used as a prior for $\lambda$.
Figure \ref{noz_area_unc} a) shows prior mean nozzle area while Figure \ref{noz_area_unc}(b) shows prior probability distribution. Comparison of prior mean with `true' nozzle area is also shown in Figure \ref{noz_area_unc}(a).
%%%%%%%%%

\begin{figure}[t!]
  \begin{center}
  \includegraphics[width=2.5in, height=2.45in] {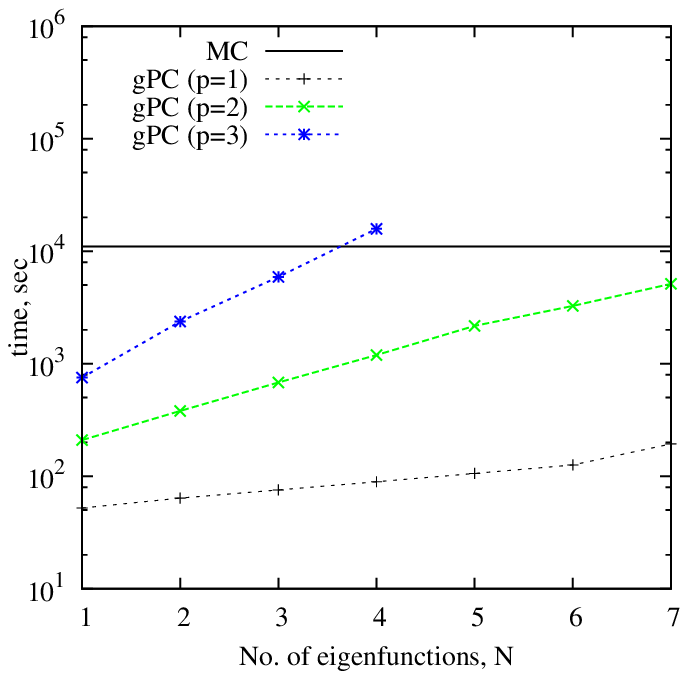}
  \end{center}
  \vspace*{-.2in}
 \caption{Figure shows cpu time required as a function of number of eigenfunctions used. Effect of the polynomial chaos order is also shown.
  }
\label{pc_mc_time}
%\end{figure}
%%%%%%%%%%
%
%\begin{figure}[tp]
  \vspace*{.2in}
  \begin{center}
  \begin {tabular}{l l}
  \subfigure []{\includegraphics[width=2.5in, height=2.45in] {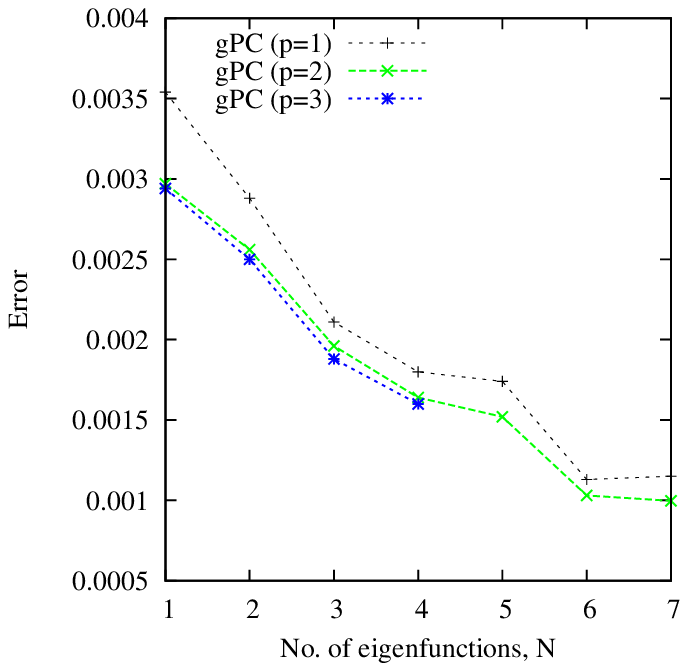}} &
    \subfigure [] {\includegraphics[width=2.5in, height=2.45in] {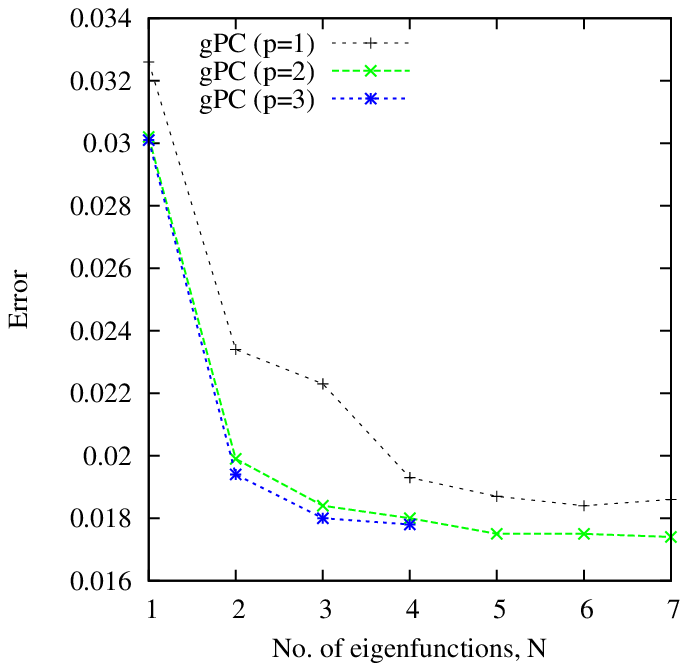}} \\
   \end{tabular}
  \end{center}
    \vspace*{-.2in}
 \caption{Figure shows (a) $L_1$-error in mean and (b) $L_1$-error in variance. Effect of the polynomial chaos order is also shown.
  }
\label{pc_mc}
\end{figure}

 The Bayesian framework is implemented using the intrusive gPC expansion for propagation of the prior uncertainty to the system response.
The prior uncertainty is projected on a Hermite polynomial chaos basis.
To investigate the trade-off between computational efficiency and accuracy, results of polynomial chaos method for uncertainty propagation are compared with the Monte Carlo simulation.
Total 10000 samples are used for the Monte Carlo method.
All the computations are performed on a desktop computer with intel i5-460 processor.
CPU time is estimated using a FORTRAN intrinsic cpu\_time routine.
Figure \ref{pc_mc_time} shows the comparison of computational time requirement as a function of number of eigenfunctions used, $N$, for different polynomial chaos order, $p$.
As can be observed from the figure, the computational cost increases polynomially with $N$.
In particular, for a system with stochasticity of order $s$, the computational cost increases as $N s^{pq}$, where $q$ is the order of non-linearity of the system and $p$ is order of the gPC basis.

Figure \ref{pc_mc} shows the $L_1$-error in the mean and the variance.
The $L_1$-error is calculated with respect to the Monte Carlo method.
With the increase in number of eigenfunctions used and the polynomial order, error in the mean and the variance reduces.
For $N=4$ and the $2^{nd}$ order polynomial chaos, method provides prediction of system response with error of the order of $10^{-3}$  in both mean and variance at $10$-times lower computational cost.
Note that the computational cost of the proposed Bayesian calibration method is dominated by the computational time requirement for solution of forward propagation problem, while the computation cost of the MCMC sampling is negligible.
Thus, the conclusions drawn from the computational cost comparison for forward propagation can be extended to solution of the inverse problem without any change.

%%%%%%%%%

\subsection{Spectral Projection-Based Bayesian Calibration}

\subsubsection{Baseline Case}

The computational cost of the proposed method is compared with the direct MCMC sampling for the Bayesian calibration.
Prior for the uncertain nozzle area is specified using the Gaussian process with the uncertain variance and the covariance length.
$IG(9.0,0.5)$ prior is used for the variance while $G(5.0,0.2)$ prior is used for the correlation length.
First four eigenfunctions are used in the spectral expansion, while $2^{nd}$ order Hermite polynomials are used as the gPC basis.
Model structural uncertainty is quantified by $IG(9.0,0.5)$ prior for the variance and $G(6.0,2.0)$ prior for the correlation length of the covariance function of the discrepancy function.
The gPC expansion coefficients of the simulator output, obtained using the intrusive Galerkin projection, are used in the likelihood function to define the posterior distribution, which is explored using the MCMC sampling.
Gaussian distribution centered on the present state is used as a proposal distribution for the Markov Chain.
Total 100000 samples are collected after rejecting the initial 10000 samples.
Results of the proposed method are compared with the direct implementation of MCMC for the Bayesian calibration, where the simulator output is used to define the likelihood.
Total 10000 samples are collected using the direct MCMC after the initial burnout period of 1000 samples.
The total CPU time required for the implementation of the Bayesian framework using the direct MCMC sampling is 12732.65 seconds.
The CPU time requirement for the proposed gPC expansion based Bayesian framework is split into the time required for the intrusive Galerkin projection of the prior uncertainty to the system response and the MCMC sampling from the posterior distribution.
For the present test case, intrusive Galerkin projection is implemented in 1194.15 seconds, while the total CPU time for the MCMC sampling is 8.48 seconds, taking 1202.63 seconds for complete implementation of the proposed stochastic spectral projection based Bayesian framework. 

\begin{figure}[tp]
  \begin{center}
  \includegraphics[width=4.0in, height=2.45in] {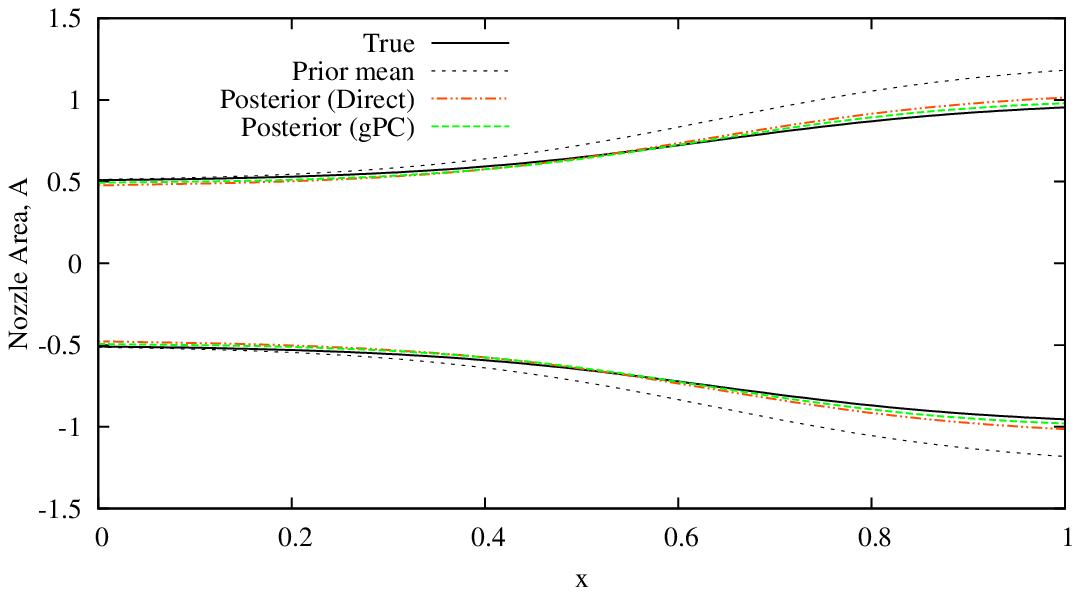}
  \end{center}
  \vspace*{-.2in}
 \caption{Comparison of the posterior mean nozzle area with the true nozzle area and prior mean. The comparison is shown for the posterior mean nozzle area obtained using implementation of the direct MCMC sampling and the proposed method.}
\label{post_noz_area_direct}
%\end{figure}
%%%%%%%%%%
%
%
%% -----------------------------------------------------------------------------
%\begin{figure}[h!]
\vspace*{.2in}
  \begin{center}
  \begin {tabular}{l l}
  \subfigure []{\includegraphics[width=2.5in, height=2.45in] {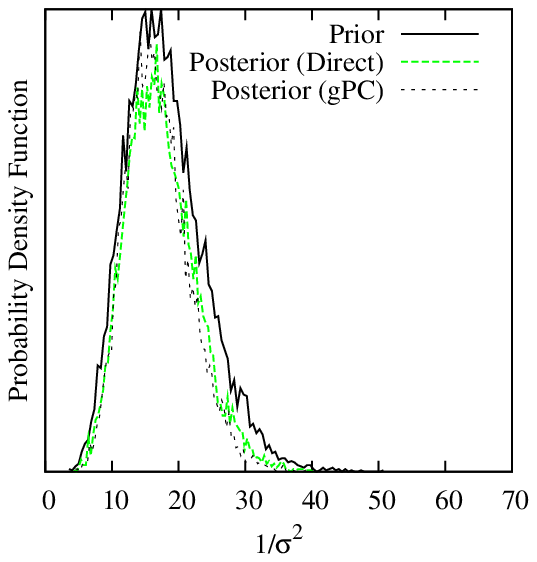}} &
    \subfigure [] {\includegraphics[width=2.5in, height=2.45in] {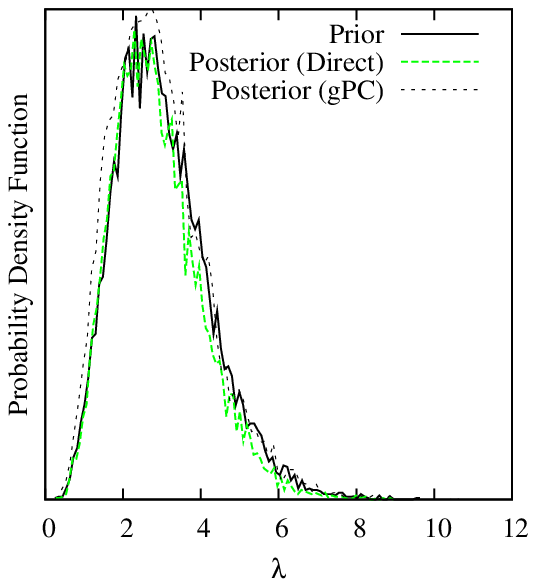}} \\
   \end{tabular}
  \end{center}
    \vspace*{-.2in}
 \caption{Figure shows comparison of posterior distribution for the hyper-parameters of the uncertain nozzle area obtained using the direct MCMC and the proposed spectral projection based method. Figure (a) shows the comparison for the variance and the Figure (b) shows the comparison for the correlation length.}
\label{post_gamma_art_direct}
\end{figure}

Figure \ref{post_noz_area_direct} shows comparison of the posterior mean nozzle area obtained using the direct MCMC and the proposed method.
The prior mean nozzle area is also shown in the figure.
The posterior mean nozzle area obtained using both the methods matches closely with the `true' nozzle area.
The comparison of the posterior distribution for the hyper-parameters of the uncertain nozzle area is shown in Figure \ref{post_gamma_art_direct}.
The probability distribution of the hyper-parameters is not updated noticeably after the Bayesian calibration, as the experimental observations of the system response are not expected to contain significant information about the covariance structure of the uncertain nozzle area.
From the figures, it may be concluded that the proposed spectral projection based Bayesian framework provide inference of the uncertain parameters with accuracy comparable to the implementation of the direct MCMC sampling at significantly lower computational cost. 

% ---------------------------------------------------------------------------
%%%%%%%%%

%----------------------------------------------------------------------------

\subsubsection{Choice of Discrepancy Function}
\begin{figure}[tp]
  \begin{center}
  \includegraphics[width=4.0in, height=2.45in] {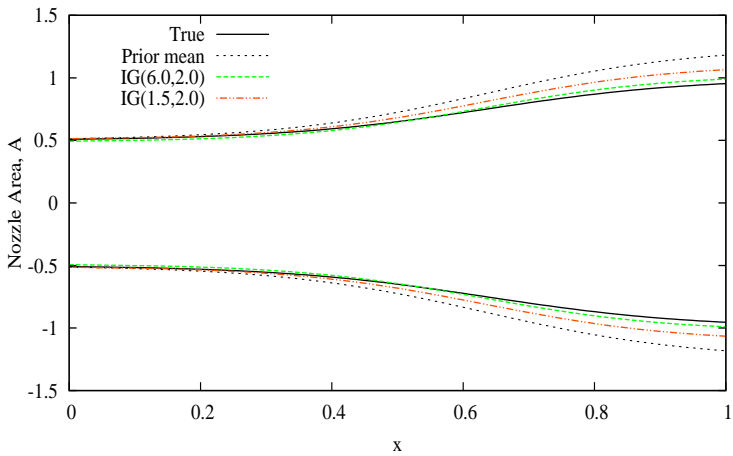}
  \end{center}
 \caption{Figure shows effect of the choice of prior for discrepancy function on posterior mean of the nozzle area.
Note that $IG(1.5,2.0)$ represent lower confidence on the simulator model as compared to the $IG(6.0,2.0)$ prior.}
\label{post_noz_area}
\end{figure}

To investigate the effect of choice of the prior for discrepancy function, Bayesian framework is implemented using different priors for variance.
Prior uncertain in the nozzle area is specified as discussed earlier.
Prior uncertainty in discrepancy function is specified using zero-mean Gaussian process with squared exponential covariance function (\ref{cov_noz_sq_exp}), where  the variance and the correlation length are uncertain hyper-parameters.
For all the test cases presented in this section, prior in correlation length is represented using $G(6.0,2.0)$, specifying correlated discrepancy function.
For variance, $IG(6.0,2.0)$ and $IG(1.5,2.0)$ priors are investigated.
To simulate model structure uncertainty, hypothetical test bed data is obtained using viscous model, whereas, simulator is defined using inviscid model.
Total 5 testbed data points are used with 1\% standard deviation.
Figure \ref{post_noz_area} shows comparison of posterior mean nozzle area with true nozzle area and the prior mean.
Posterior mean for $IG(6.0,2.0)$ prior matches closely with the true nozzle area, whereas, posterior mean nozzle area for $IG(1.5,2.0)$ prior deviates from the true nozzle area.
%%%%%%%%%
Figure \ref{post_noz_area} demonstrates the significant impact of the prior model structural uncertainty on the inference of the uncertain parameters.
Confidence on the available model is specified through the prior on the variance.
In the case of high confidence on the simulator model, specified through the high value of $\alpha$ for $IG(\alpha,\beta)$ prior, significant amount of the information provided by the data is used to update the parameter uncertainty.
If the low confidence prior is specified for the model through the lower value of $\alpha$ for $IG(\alpha,\beta)$ prior, the calibration process uses significant information provided by the data to improve confidence on the simulator model, whereas, less information is used for updating the uncertain parameters.
Note that in the case of very high confident prior on the model structure (approaching the scenario of no model structural uncertainty), calibration process attributes any remnant error in the model to the parameters.
Thus, the present authors propose to avoid priors that specify either low confidence or very high confidence on the model structure.
In the remaining test cases presented in this paper, $IG(6.0,2.0)$ prior is used for the variance of the discrepancy function.

\subsubsection{Effect of Simulator Model Error}
\begin{figure}[tp]
  \begin{center}
  \includegraphics[width=4.0in, height=2.45in] {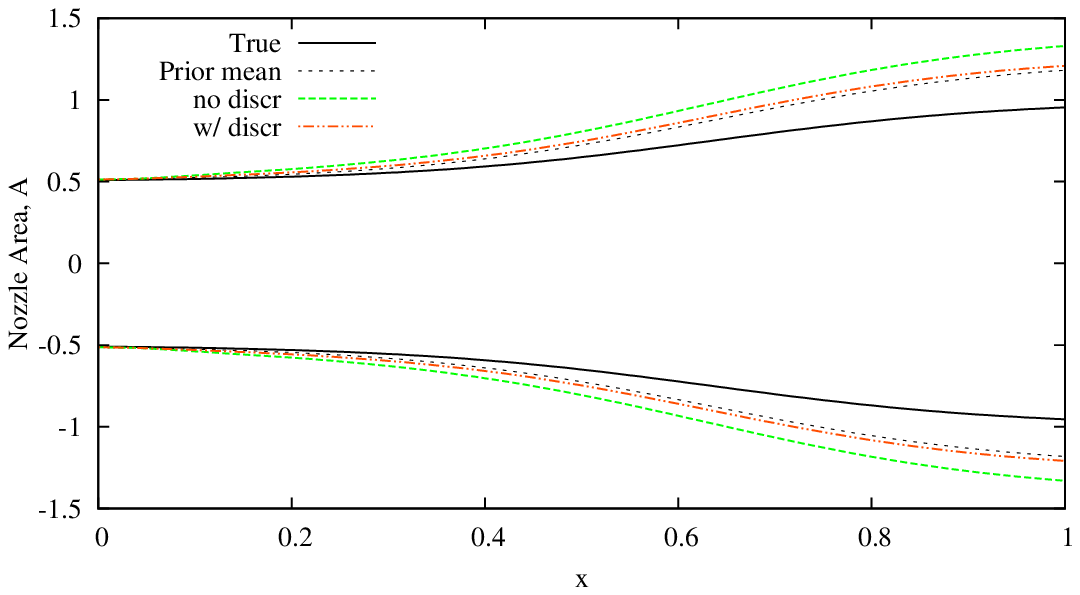}
  \end{center}
    \vspace*{-.2in}
 \caption{Figure shows the effect of error in the simulator model on the posterior nozzle area. Results are obtained for artificially introduced discrepancy in the simulator model. The comparison is shown for the Bayesian calibration without using the discrepancy function (no discr) and with using the discrepancy function (w/ discr).}
\label{noz_area_art_discr}
%\end{figure}
%%%%%%%%%%
%% -----------------------------------------------------------------------------
%% -----------------------------------------------------------------------------
%\begin{figure}[h!]
  \vspace*{.2in}
  \begin{center}
  \begin {tabular}{l l}
  \subfigure []{\includegraphics[width=2.5in, height=2.45in] {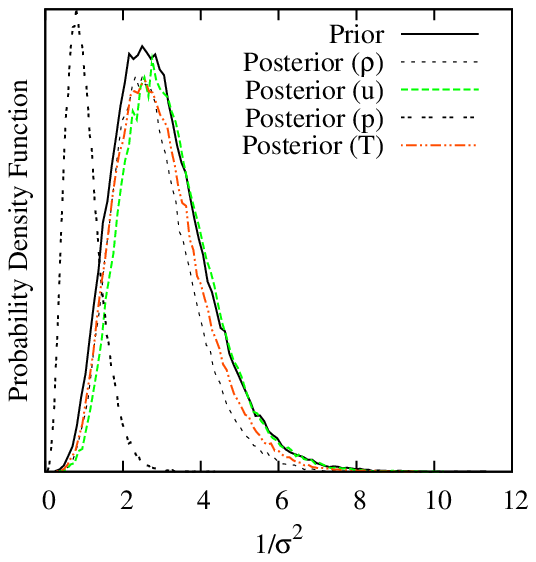}} &
    \subfigure [] {\includegraphics[width=2.5in, height=2.45in] {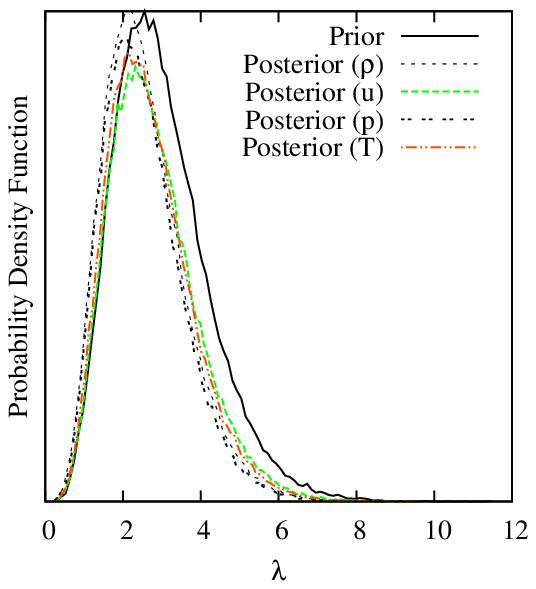}} \\
   \end{tabular}
  \end{center}
  \vspace*{-.2in}
 \caption{Figure shows comparison of prior and posterior distribution for a) $\sigma^2_\delta$ and b) $\lambda_\delta$ in presence of error in the simulator model. Results are obtained for artificially introduced discrepancy in the simulator model.
 }
\label{post_gamma_art_discr}
\end{figure}

To investigate efficacy of the calibration process in presence of error in simulator model, artificial discrepancy is introduced in the simulator model by multiplying the gPC expansion coefficients of the static pressure by 1.5.
Modified chaos coefficients are used in the Bayesian framework.
The framework is implemented using following two test cases: (1) simulator model without taking into account discrepancy function (which signifies full confidence on the simulator model), and (2) $G(6.0,2.0)$ and $IG(6.0,2.0)$ priors for correlation length and variance of the discrepancy function ($\lambda_\delta$ and $\sigma^2_\delta$ respectively). 
Figure \ref{noz_area_art_discr} shows comparison of posterior nozzle area with the true nozzle area.
For calibration without considering discrepancy function, Bayesian framework assumes simulator to be the `true' representation of the physics.
The Bayesian calibration method attributes all the difference between the test bed data and the simulator prediction to the uncertain parameters.
Thus, resultant posterior mean deviates significantly from the true nozzle area. However, when discrepancy function is considered in the Bayesian calibration, no significant update is observed in the nozzle area.

Figure \ref{post_gamma_art_discr} shows comparison of the prior and the posterior probability distribution of $\sigma^2_\delta$ and $\lambda_\delta$.
The posterior probability distribution of $1/{\sigma^2_\delta}$ for the static pressure is shifted towards left, indicating the lower posterior expected value of $1/{\sigma^2_\delta}$ as compared to the prior.
However, no significant change is observed for the probability distribution of $\lambda_\delta$.
Note that the prior $G(6.0,2.0)$ indicated correlated discrepancy in the simulator model.
Thus, the posterior probability distribution of the discrepancy function for the static pressure indicates correlated discrepancy in the simulator model with the high expected value of $\sigma^2_\delta$.
As the error in the experimental observations enters the Bayesian framework as the uncorrelated uncertainty, the highly correlated discrepancy is expected to result from the error in the simulator model.
Thus, the posterior distribution of the discrepancy function signals the need for verification and validation of the simulator, particularly subroutines affecting the static pressure.
% -----------------------------------------------------------------------------
%%%%%%%%%

\subsubsection{Effect of Erroneous Observations}
\begin{figure}[tp]
  \begin{center}
  \includegraphics[width=4.0in, height=2.45in] {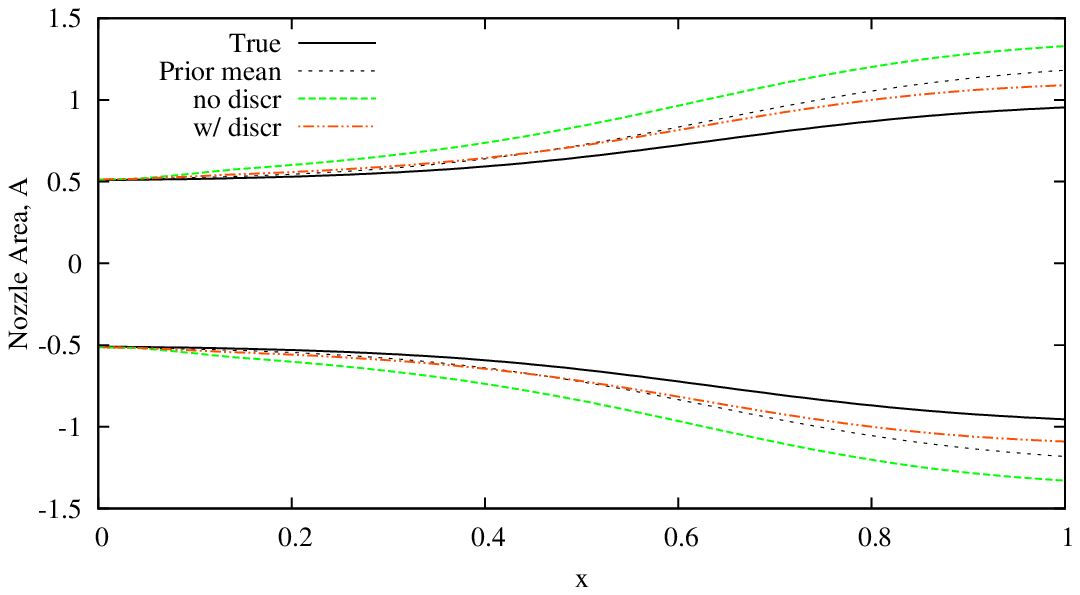}
  \end{center}
   \vspace*{-.2in}
 \caption{Figure shows the effect of erroneous experimental observations on the posterior nozzle area. Results are obtained for artificially introduced errors in the experimental observations.} %The comparison is shown for the Bayesian calibration without using the discrepancy function (no discr) and with using the discrepancy function (w/ discr).}
\label{noz_area_art_discr_expt}
%\end{figure}
%%%%%%%%%%
%% ------------------------------------------------------------------------------
%% -----------------------------------------------------------------------------
%\begin{figure}[h!]
  \vspace*{.2in}
  \begin{center}
  \begin {tabular}{l l}
  \subfigure []{\includegraphics[width=2.5in, height=2.45in] {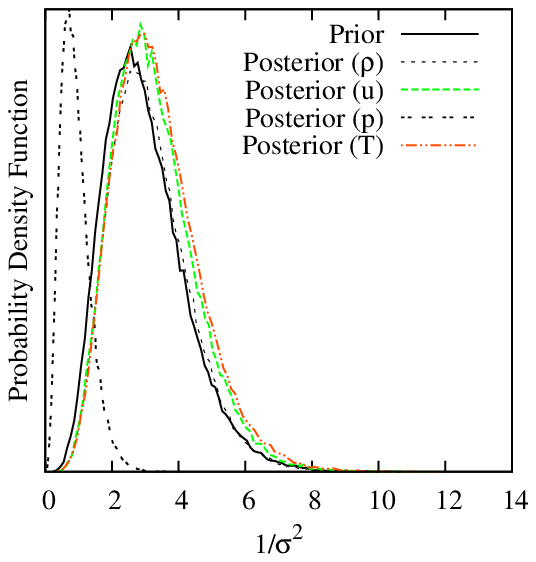}} &
    \subfigure [] {\includegraphics[width=2.5in, height=2.45in] {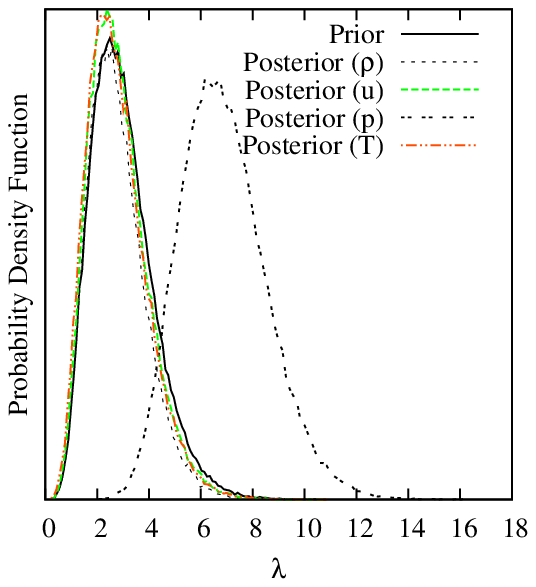}} \\
   \end{tabular}
  \end{center}
  \vspace*{-.2in}
 \caption{Figure shows comparison of prior and posterior distribution for a) $\sigma^2_\delta$ and b) $\lambda_\delta$ in presence of erroneous experimental observations. }%Results are obtained for artificially introduced error in the experimental observations.}
\label{post_gamma_art_discr_expt}
\end{figure}

To investigate effect of the erroneous experimental observations on the Bayesian framework, proposed method is implemented with artificially introduced discrepancy in test bed data.
The methodology is demonstrated by multiplying 1.5 to static pressure data.
Figure \ref{noz_area_art_discr_expt} shows resultant posterior mean nozzle area.
As observed in case of artificially introduced discrepancy in the simulator, significant deviation is observed between posterior mean and true nozzle area when discrepancy function is not considered in the formulation.
However, when discrepancy function is considered, no significant change is observed in the nozzle area.
% -----------------------------------------------------------------------------
%%%%%%%%%
% -----------------------------------------------------------------------------

Figure \ref{post_gamma_art_discr_expt} shows posterior probability distribution for $\sigma^2_\delta$ and $\lambda_\delta$.
Posterior distribution of $1/\sigma^2_\delta$ for static pressure has moved towards left, indicating the high posterior discrepancy.
Posterior probability distribution of $\lambda_\delta$ for static pressure is shifted towards right, indicating highly uncorrelated discrepancy.
Since the uncertainty due to unknown or poorly known physics is expected to result in the correlated discrepancy, uncorrelated discrepancy may be attributed to the experimental observations.
Thus the posterior distribution indicates need for the review of the experimental observations.

\subsubsection{Summary}
The numerical test cases presented in this paper have demonstrated ability of the proposed Bayesian framework to infer spatially/temporally varying uncertain parameters in the presence of model structural uncertainty.
In addition to the inference of the uncertain parameters, the proposed Bayesian framework provides useful insight into the credibility of the simulator model.
The posterior probability distributions of $\sigma^2_\delta$ and $\lambda_\delta$, through hyper-parameters $\alpha$ and $\beta$, are identified as indicators of the veracity and validity of the simulator model.
Based on the posterior values of the hyper-parameters $\alpha$ and $\beta$ for the posterior probability distributions of $\sigma^2_\delta$ ($\alpha_{\sigma_\delta}$ and $\beta_{\sigma_\delta}$) and $\lambda_\delta$ ($\alpha_{\lambda_\delta}$ and $\beta_{\lambda_\delta}$), authors provide following guidelines:
\begin{enumerate}
\item If the posterior $\alpha_{\sigma_\delta}$ is greater than the prior $\alpha_{\sigma_\delta}$, then the simulator model should be accepted with improved confidence.
\item $\alpha_{\sigma_\delta} < 1$ : The posterior distribution does not have mode resulting in maximum probability for $\frac{1}{\sigma^2_\delta} \rightarrow 0$. For such posterior, calibrated simulator model should be rejected.
\begin{itemize}
\item If mean and mode of posterior distribution of $\lambda_\delta$ indicate strong correlation for discrepancy function (typically $\alpha_\lambda < 1$ or $\beta_\lambda>\alpha_\lambda$), rigorous verification and validation process is advised with a focus on subsystem models that predict system responses for which $\alpha_{\sigma_\delta} < 1$.
\item If mean and mode of posterior distribution of $\lambda_\delta$ indicate very weak correlation for discrepancy function (typically $\alpha_\lambda> 1$ or $\beta_\lambda << \alpha$), review of experimental observations is advised.
\item For all the other cases, rigorous verification and validation of simulator model is advised with a note on review of experimental observations.
\end{itemize}
\item $\alpha_{\sigma_\delta} > 1$ and $\beta_{\sigma_\delta} > \alpha_{\sigma_\delta}$: Calibrated simulator can be used, however, high uncertainty in prediction of the simulator response should be expected. As per discussion void point (2), verification and validation of simulator model and a review of experimental observations is advised.
\item $\alpha_{\sigma_\delta} >>1$ and $\beta_{\sigma_\delta}  < \alpha_{\sigma_\delta}$: Calibrated simulator can be used with high confidence.
\end{enumerate}

\section{Concluding Remarks}

This paper has demonstrated computational efficiency of a gPC based Bayesian framework for calibration of a large scale system simulator.
The proposed framework has extended the established method to the priors with uncertain hyper-parameters.
Efficacy of the proposed Bayesian framework is demonstrated for calibration of a quasi-one-dimensional divergent nozzle flow simulator.
Ability of the method to infer spatially/temporally varying uncertain parameters is shown using the update of the nozzle area.
The proposed method has provided accurate inference of nozzle area at one-tenth of a computational cost as compared to the direct implementation of the Bayesian framework.
Hyper-parameters of the posterior distribution of model structure uncertainty are identified that provide information about veracity and validity of the computer simulator.
Based on the hyper-parameters, guidelines have been provided for acceptability of the simulator model.  
Although demonstrated for a specific set of priors, the proposed method is generic in nature and can admit arbitrary priors.
However, depending on the gPC basis used, higher order polynomials may be required for the satisfactory spectral approximation of the prior, incurring comparatively higher computational cost on the Bayesian framework.

\section*{Acknowledgements}
This work was supported by Basic Science Research Program through the National Research Foundation of Korea (NRF) funded by the Ministry of Education, Science and Technology (2010-0025484).

% --------------- References ----------------------
\bibliographystyle{gIPE}
\bibliography{Refer_ipse}

\begin{thebibliography}{47}
\providecommand{\natexlab}[1]{#1}

\bibitem[1]{Mehta91}
U. Mehta, {\itshape Some aspects of uncertainty in computational fluid dynamics
  results}, J. of Fluid Engg. 113 (1991), pp. 538--543.

\bibitem[2]{Mehta96}
---{}---{}---, {\itshape Guide to credible computer simulations of fluid
  flows}, J. of Prop. and Power 12 (1996), pp. 940--948.

\bibitem[3]{OreskesScience94}
N. Oreskes, K. Shrader-Frechett, and K. Belitz, {\itshape Verification,
  validation and confirmation of numerical models in earth sciences}, Science
  263 (1994), pp. 641--647.

\bibitem[4]{Mehta98}
U. Mehta, {\itshape Credible computational fluid dynamics simulations}, AIAA J.
  36 (1998), pp. 665--667.

\bibitem[5]{OberkampfRESS02}
W. Oberkampf, S. DeLand, B. Rutherford, K. Diegert, and K. Alvin, {\itshape
  Error and uncertainty in modeling and simulation}, Rel. Engg. and Syst.
  Safety 75 (2002), pp. 335--357.

\bibitem[6]{ThunPhD}
D. Thunnissen, {\itshape Propagating and mitigating uncertainty in the design
  of complex multidisciplinary systems}, Ph.D. diss., California Institute of
  Technology, California, 2004.

\bibitem[7]{cheung11}
S.H. Cheung, T.A. Oliver, E.E. Prudencio, S. Prudhomme, and R.D. Moser,
  {\itshape Bayesian uncertainty analysis with applications to turbulence
  modeling}, Rel. Engg. and Syst. Safety 96 (2011), pp. 1137--1149.

\bibitem[8]{TrucanoRESS06}
T. Trucano, L. Swiler, T. Igusa, W. Oberkampf, and M. Pilch, {\itshape
  Calibration, validation, and sensitivity analysis: {W}hat's what}, Rel. Engg.
  and Syst. Safety 91 (2006), pp. 1331--1357.

\bibitem[9]{Glim99}
J. Glimm and D. Sharp, {\itshape Prediction and the quantification of
  uncertainty}, Physica D 133 (1999), pp. 152--170.

\bibitem[10]{KennedyJRSS01}
M. Kennedy and A. O'Hagan, {\itshape Bayesian calibration of computer models},
  J. of the Royal Stat. Soc. Series B (Stat. Method.) 63 (2001), pp. 425--464.

\bibitem[11]{HigdonJSC04}
D. Higdon, M. Kennedy, J. Cavendish, J. Cafeo, and R. Ryne, {\itshape Combining
  field data and computer simulations for calibration and prediction}, SIAM J.
  of Sci. Comp. 26 (2005), pp. 448--446.

\bibitem[12]{Goldstein04}
M. Goldstein and J. Rougier, {\itshape Probabilistic formulations for
  transferring inferences from mathematical models to physical systems}, SIAM
  J. of Sci. Comp. 26 (2005), pp. 467--487.

\bibitem[13]{Bayarri_Tech07}
M. Bayarri, J. Berger, R. Paulo, J. Sacks, J. Cafeo, J. Cavendish, C. Lin, and
  J. Tu, {\itshape A framework for validation of computer models},
  Technometrics 49 (2007), pp. 138--153.

\bibitem[14]{higdon_jasa08}
D. Higdon, J. Gattiker, B. Williams, and M. Rightley, {\itshape Computer model
  calibration using high-dimensional output}, J. of the Amer. Stat. Assoc. 103
  (2008), pp. 570--583.

\bibitem[15]{KellyRESS09}
D. Kelly and C. Smith, {\itshape Bayesian inference in probabilistic risk
  assessment - the current state of the art}, Rel. Engg. and Sys. Safety
  (2009), pp. 628--643.

\bibitem[16]{goldstein_jspi09}
M. Goldstein and J. Rougier, {\itshape Reified {B}ayesian modelling and
  inference for physical systems}, J. of Stat. Planning and Inf. 139 (2009),
  pp. 1221--1239.

\bibitem[17]{BesagSS95}
J. Besag, P. Green, D. Higdon, and K. Mengersen, {\itshape Bayesian computation
  and stochastic systems}, Stat. Sci. 10 (1995), pp. 3--41.

\bibitem[18]{Gamerman}
D. Gamerman and H. Lopes {\itshape Markov {C}hain {M}onte {C}arlo: {S}tochastic
  {S}imulation for {B}ayesian {I}nference},    Chapman and Hall/CRC, Boca
  Raton, 2006.

\bibitem[19]{MarzoukJCP07}
Y. Marzouk and H. Najm, {\itshape Stochastic spectral methods for efficient
  {B}ayesian solution of inverse problems}, J. of Comp. Phys.  (2007), pp.
  560--586.

\bibitem[20]{Walters02}
R. Walters and L. Huyse, {\itshape Uncertainty analysis for fluid mechanics
  with applications}, in {\itshape NASA/CR-2002-211449}, 2002.

\bibitem[21]{WienerAJM38}
N. Wiener, {\itshape The homogeneous chaos}, Amer. J. of Math. 60 (1938), pp.
  897--936.

\bibitem[22]{Wiener}
N. Wiener {\itshape Nonlinear {P}roblems in {R}andom {T}heory},    John Wiley
  \& Sons, New York, 1958.

\bibitem[23]{CameronAM47}
R. Cameron and W. Martin, {\itshape The orthogonal development of non-linear
  functionals in series of {F}ourier-{H}ermite functionals}, The Annals of
  Math. 48 (1947), pp. 385--392.

\bibitem[24]{MeechamJFM68}
W. Meecham and D. Jeng, {\itshape Use of the {W}iener-{H}ermite expansion for
  nearly normal turbulence}, J. of Fluid Mech. 32 (1968), pp. 225--249.

\bibitem[25]{OrszagPF67}
S. Orszag and L. Bissonnette, {\itshape Dynamical properties of truncated
  {W}iener-{H}ermite expansions}, Phys. of Fluids 10 (1967), pp. 260--263.

\bibitem[26]{ChorinJFM74}
A. Chorin, {\itshape Gaussian fields and random flow}, J. of Fluid Mech. 85
  (1974), pp. 325--347.

\bibitem[27]{Ghanem91}
R. Ghanem and P. Spanos, {\itshape Spectral stochastic finite-element
  formulation for reliability analysis}, J. of Engg. Mech. 117 (1991), pp.
  2351--2372.

\bibitem[28]{GhanemPhyD99}
R. Ghanem and J. Red-Horse, {\itshape Propagation of probabilistic uncertainty
  in complex physical systems using a stochastic finite element approach},
  Physica D 133 (1999), pp. 137--144.

\bibitem[29]{Ghanem}
R. Ghanem and P. Spanos {\itshape Stochastic {F}inite {E}lements: {A}
  {S}pectral {A}pproach},    Dover Publications, 2003.

\bibitem[30]{Knio_FDR_2006}
O. Knio and O. Maitre, {\itshape Uncertainty propagation in {CFD} using
  polynomial chaos decomposition}, Fluid Dyn. Res. 38 (2006), pp. 616--640.

\bibitem[31]{Maitre_2001}
O. Maitre, O. Knio, H. Najm, and R. Ghanem, {\itshape A stochastic projection
  method for fluid flow}, J. of Comp. Phys. 173 (2001), pp. 481--511.

\bibitem[32]{Xiu_SJSC_2004}
D. Xiu and G. Karniadakis, {\itshape The {W}einer-{A}skey polynomial chaos for
  stochastic differential equations}, SIAM J. of Sci. Comp. 24 (2002), pp.
  619--644.

\bibitem[33]{XiuJCP03}
---{}---{}---, {\itshape Modeling uncertainty in flow simulations via
  generalized polynomial chaos}, J. of Comp. Phys. 187 (2003), pp. 137--167.

\bibitem[34]{Koekoek}
R. Koekoek and R. Swarttouw {\itshape The Askey-scheme of hypergeometric
  orthogonal polynomials and its q-analogue},    Department of Technical
  Mathematics and Informatics, Report no. 98-17, Delft University of
  Technology, 1998.

\bibitem[35]{Lucor03}
D. Lucor, D. Xiu, C. Su, and G. Karniadakis, {\itshape Predictability and
  uncertainty in {CFD}}, Int. J. for Num. Meth. in Fluids 43 (2003), pp.
  483--505.

\bibitem[36]{Mathelin04}
L. Mathelin, M. Hussaini, T. Zang, and F. Bataille, {\itshape Uncertainty
  propagation for a turbulent, compressible nozzle flow using stochastic
  methods}, AIAA J. 42 (2004), pp. 1669--1676.

\bibitem[37]{Narayanan04}
V. Narayanan and N. Zabaras, {\itshape Stochastic inverse heat conduction using
  spectral approach}, Int. J. for Num. Methods in Engg.  (2004), pp.
  1569--1593.

\bibitem[38]{Najm_afm}
H. Najm, {\itshape Uncertainty quantification and polynomial chaos techniques
  in computational fluid dynamics}, Ann. Rev. of Fluid Mech. 41 (2009), pp.
  35--52.

\bibitem[39]{MarzoukJCP09}
Y. Marzouk and H. Najm, {\itshape Dimensionality reduction and polynomial chaos
  acceleration of {B}ayesian inference in inverse problems}, J. of Comp. Phys.
  (2009), pp. 1862--1902.

\bibitem[40]{TagadeIDETC11}
P. Tagade and H.L. Choi, {\itshape A polynomial chaos based {B}ayesian
  inference method with uncertain hyperparameters}, in {\itshape ASME
  International Design Engineering Technical Conference and Computers and
  Information in Engineering Conference, Washington, DC, USA}, 2011.

\bibitem[41]{SacksSS89}
J. Sacks, W. Welch, T. Mitchell, and H. Wynn, {\itshape Design and analysis of
  computer experiments}, Stat. Sci. 4 (1989), pp. 409--423.

\bibitem[42]{PauloAS05}
R. Paulo, {\itshape Default priors for {G}aussian processes}, The Ann. of Stat.
  33 (2005), pp. 556--582.

\bibitem[43]{O'HaganRESS06}
A. O’Hagan, {\itshape Bayesian analysis of computer code outputs: {A}
  tutorial}, Rel. Engg. and Syst. Safety 91 (2006), pp. 1290--1300.

\bibitem[44]{MetropolisJCP53}
N. Metropolis, A. Rosenbluth, M. Rosenbluth, A. Teller, and E. Teller,
  {\itshape Equation of state calculations by fast computing machines}, The J.
  of Chem. Phys. 21 (1953), pp. 1087--1092.

\bibitem[45]{HastingsBio70}
W. Hastings, {\itshape Monte {C}arlo sampling methods using {M}arkov chains and
  their applications}, Biometrika 57 (1970), pp. 97--109.

\bibitem[46]{fie_09}
A. Chakrabarti and S. Martha, {\itshape Approximate solutions of {F}redholm
  integral equations of the second kind}, App. Math. and Comp. 211 (2009), pp.
  459--466.

\bibitem[47]{Huang_ijnme01}
S. Huang, S. Quek, and K. Phoon, {\itshape Convergence study of the truncated
  {K}arhunen–{L}oeve expansion for simulation of stochastic processes}, Int.
  J. of Num. Methods in Engg. 52 (2001), pp. 1029--1043.

\end{thebibliography}
% --------------------------------------------------

\label{lastpage}

\end{document}